\documentclass[conference]{IEEEtran}
\IEEEoverridecommandlockouts

\usepackage[font=footnotesize]{caption}
\usepackage{subfig}

\usepackage{graphicx}
\usepackage{wrapfig}
\usepackage{lipsum}

\usepackage{tabularx}
\usepackage{amssymb}
\usepackage[usenames, dvipsnames]{color}
\usepackage{algorithm}
\usepackage{algpseudocode}
\usepackage{etoolbox}

\usepackage{tikz}
\newcommand\copyrighttext{\footnotesize \textcopyright 2018 IEEE. Personal use of this material is permitted. Permission from IEEE must be obtained for all other uses, in any current or future media, including reprinting/republishing this material for advertising or promotional purposes, creating new collective works, for resale or redistribution to servers or lists, or reuse of any copyrighted component of this work in other works. \\
Accepted to be Published in: Proceedings of the 2018 IEEE 24th International Symposium on On-Line Testing And Robust System Design (IOLTS), July 2-4, 2018, Platja d’Aro, Costa Brava, Spain}
\newcommand\copyrightnotice{%
\begin{tikzpicture}[remember picture,overlay]
\node[anchor=south,yshift=10pt] at (current page.south) {\fbox{\parbox{\dimexpr\textwidth-\fboxsep-\fboxrule\relax}{\copyrighttext}}};
\end{tikzpicture}%
}

\renewcommand{\baselinestretch}{0.96}

\begin{document}
\title{Benchmarking the Capabilities and Limitations of SAT Solvers in Defeating Obfuscation Schemes}

\author{\IEEEauthorblockN{Shervin Roshanisefat, Harshith K. Thirumala, Kris Gaj, Houman Homayoun, Avesta Sasan}
\IEEEauthorblockA{
Department of Electrical and Computer Engineering, George Mason University, Fairfax, U.S.A.\\
e-mail: \{sroshani, hthiruma, kgaj, hhomayou, asasan\}@gmu.edu
}}

\makeatletter
\patchcmd{\@maketitle}
  {\addvspace{0.5\baselineskip}\egroup}
  {\addvspace{-1\baselineskip}\egroup}
  {}
  {}

\maketitle

\begin{abstract}
In this paper, we investigate the strength of six different SAT solvers in attacking various obfuscation schemes. Our investigation revealed that Glucose and Lingeling SAT solvers are generally suited for attacking small-to-midsize obfuscated circuits, while the MapleGlucose, if the system is not memory bound, is best suited for attacking mid-to-difficult obfuscation methods. Our experimental result indicates that when dealing with extremely large circuits and very difficult obfuscation problems, the SAT solver may be memory bound, and Lingeling, for having the most memory efficient implementation, is the best suited solver for such problems. Additionally, our investigation revealed that SAT solver execution times may vary widely across different SAT solvers. Hence, when testing the hardness of an obfuscation methods, although the increase in difficulty could be verified by one SAT solver, the pace of increase in difficulty is dependent on the choice of a SAT solver.
\end{abstract}


\IEEEpeerreviewmaketitle
\copyrightnotice

\vspace{-15pt}
\section{Introduction}
Cost of building a new semiconductor fab was estimated to be US \$5.0 billion in 2015, with large recurring maintenance costs \cite{DIGITIMES}\cite{6926108}, and sharply increases as technology migrates to smaller nodes. Due to the high cost of building, operating, managing, and maintaining state-of-the-art silicon manufacturing facilities, many major U.S. high-tech companies have been always fabless or went fabless in recent years. To reduce the fabrication cost, and for economic feasibility, most of the manufacturing and fabrication is pushed offshore \cite{DIGITIMES}. However, many offshore fabrication fabs are untrusted, which has raised concern over potential attackers that include the manufacturers, with an intimate knowledge of the fabrication process, the ability to modify and expand the design prior to production, and an unavoidable access to the fabricated chips during testing. Hence, fabrication in untrusted fabs has introduced multiple forms of security threats from supply chain including that of overproduction, Trojan insertion, Reverse Engineering (RE), Intellectual Property (IP) theft, and counterfeiting \cite{6926108}.

One of the solutions explored by researchers to resist such hardware security threats is through the application of logic obfuscation. Logic obfuscation is the process of hiding the functionality of an IP by building ambiguity or by implementing post manufacturing means of control and programmability into its netlist. Gate camouflaging and circuit locking are two of the widely explored obfuscation mechanisms \cite{6881480}\cite{7128395}\cite{7479225}. A camouflaged gate is a gate that after RE (by means of delayering and lithography) could be mapped to any of possible set of gates or may look like one logic gate, however functionally perform as a different gate. On the other hand, in locking solutions, the functionality of a circuit is locked using a number of key inputs such that only when the correct key is applied, the circuit resumes its expected functionality. Otherwise, the correct function is hidden between many of the $2^K$ ($K$ being the number of keys) different possibilities of the circuit. The claim raised by such obfuscation scheme was that to break the obfuscation, the adversaries need to try a large number of inputs and key combinations to extract the correct key, and the difficulty of this process increases exponentially with the number of keys. Hence, if enough gates are obfuscated, the adversary needs a considerable amount of time (claimed as years to decades) to deobfuscate the circuit.

The validity and strength of logic obfuscation to defend the IP against adversaries in the manufacturing supply chain was seriously challenged as researchers demonstrated that the satisfiability (SAT) solvers could break the obfuscation in a matter of minutes as opposed to the promised claim of years and decades \cite{7140252}\cite{el2015integrated}. This redirected the attention of the researchers to find harder obfuscation schemes that are more resilient to SAT attacks. SARLock and Anti-SAT \cite{7495588}\cite{xie2016mitigating} obfuscation methods were proposed for this purpose, however further research proved that these obfuscation techniques are prone to a removal using Signal Probability Skew (SPS) attacks \cite{7858346}, leaving the problem of finding a SAT and SPS resilient obfuscation unresolved. 

Today, many off-the-shelf SAT solvers with various capabilities are freely and openly available, and each year many new and more capable solutions are being developed. Some of the most efficient and most powerful SAT solvers are the winners of the International SAT competition \cite{balyo2017proceedings}, where solvers participating in the competition are required to test the performance of the proposed SAT solvers on a large number of SAT problems in various categories. Different SAT solutions have offered widely varying performance dealing with various SAT problems, illustrating that the choice of a SAT solver and its underlying features could have a significant impact on the solver's success and also on the time it takes to solve a specific problem. In addition, different SAT solvers require various amounts of memory resources, and assuming powerful SAT solvers may perform extremely poorly or fail to find the solution for larger benchmarks, even if they may outperform other SAT solvers for solving a large number of small SAT problems. In this work, we investigate the limitations and capabilities of different classes of SAT solvers when specifically dealing with the problem of circuit obfuscation.

The contribution of this work to the hardware security community is as follows: (1) to the best of our knowledge, this work is the first attempt in benchmarking the capabilities of SAT solvers when specifically dealing with hardware obfuscation problem. It provides insights on capabilities and limitation of different classes of SAT solvers, helping the researchers in choosing the most able SAT solver for evaluating the effectiveness and hardness of their proposed obfuscation solutions and prevents researchers from generalizing the failing result of a poor choice of SAT solver solution, to all SAT solvers; (2) this work captures and summarizes the best approach for converting various obfuscation schemes into SAT solvable problems and compares the hardness of several of the previously proposed obfuscation techniques across different classes of solvers. 

\section{Preparing Obfuscated Netlists for SAT Attack}

\subsection{Converting Obfuscated Gates to Key-Programmable Gates}
A SAT solver takes a Boolean function in Conjunctive Normal Form (CNF) as input and finds a valid assignment for input variables to satisfy the function. To attack an obfuscated netlist using a SAT solver, a working copy of the chip and its obfuscated netlist is required. The adversary can acquire the working chip after it is unlocked by the manufacturer and shipped to the market and could gain access to the obfuscated netlist by means of RE. In case of supply chain adversary, the obfuscated netlist is readily available to the attacker. Then, the obfuscated netlist should be transformed into a circuit SAT problem. This process is explained next:
 
Let us refer to the functional black-box copy of the obfuscated circuit as $C_F$. The $C_F$ is used to find the correct output for any given input. When using $K$ keys, random assignment of key could create at most $2^K$ instances of a circuit. Similar argument applies to camouflaged cells, where each of $K$ camouflaged gates could assume one of the $M$ different possibilities (for simplicity, let us consider $M=2$). Let us denote obfuscation scheme obtained by means of using K keys or obfuscated gates by K-obfuscation. A circuit $C$ with $N_X$ inputs that is subjected to K-camouflaging could be represented with an equivalent $C_K$ circuit with $N_X+K$ inputs. Let us denote the circuit $C$ with input $X$ and output $Y$ by $C(X,Y)$ and its K-obfuscated netlist by $C(X,K,Y)$. If the correct set of keys $\hat{K}=(k_0, k_1, ...,k_{K-1})$ is applied to the obfuscated circuit, for every input the obfuscated circuit reduces to the original circuit $C(X,\hat{K},Y_K) \triangleq C(X,Y)$.

For a SAT attack the key signals in $C(X,K,Y)$ should be available as input. Hence, obfuscation cells should be represented as \textit{Key-Programmable Gate} (KPG), where insertion of the correct key converts them to the correct gate. The cells used for obfuscation could be divided into two categories: (1) key-controlled gates \cite{5604160}\cite{2018arXiv180409162R} in which the key is an input signal (e.g. XOR, MUX based obfuscation). (2) keyless-gates \cite{mardani2018lutlcok}\cite{li2017provably} where functionality is hidden in the ambiguous structure or by use of internal memory elements (e.g. camouflaged gates and LUTs). When using key-controlled gates, the key is stored in an internal memory or a burned fuse. Hence, in a reverse-engineered netlist the key inputs could be identified by tracking their connectivity to memory/fuse elements. To prepare the $C(X,K,Y)$ netlist, the memory/fuse element is removed and key inputs are connected to input port(s).

\begin{wrapfigure}{r}{0.32\linewidth}
    \includegraphics[width=\linewidth]{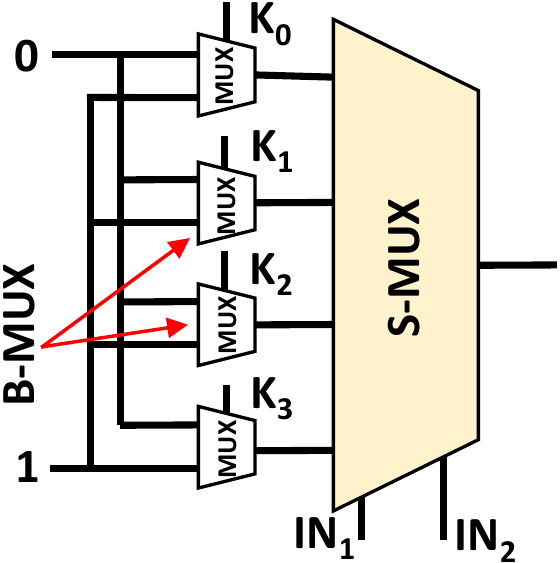}
    \caption{Converting a LUT to a KPG}
    \label{fig:lut}
\end{wrapfigure}

When using keyless-gates, the gate has to be transformed to a key-programmable gate before invoking a SAT attack. For a $L$-input LUT, the number of functional possibilities is $2^{2^L}$. To build a KPG for a LUT, the circuit illustrated in Fig. \ref{fig:lut} is deployed. The inputs to the LUT are connected to the select lines of the S-MUX and keys are the select lines of B-MUXes. Then, each key is connected to an input port adding $2^L$ keys to the $C(X,K,Y)$.

A camouflaged cell relies on hiding the gate functionality by keeping the structure of several gates similar. Even in the best camouflaging cells, the number of gate possibilities is limited and it could be treated similarly to programmable cells, where the camouflaged cell is replaced by a MUX and each of the gate possibilities is fed to a different input of the MUX, while using the select lines of the MUX as key inputs that are routed to the input pins of the $C(X,K,Y)$.

\subsection{Converting an Obfuscated Circuit Into a SAT Problem} \label{SATC_construction}
Before invoking the SAT solver, every key input combination is considered as a candidate key. Let's denote the \textit{Set of Candidate Keys} by SCK. If we can find an input $x_d$, and two distinct key values $K_1$ and $K_2$ in SCK such that $C(x_d, K_1, Y_1) \ne C(x_d, K_2, Y_2)$, the input $x_d$ is denoted as a \textit{Discriminating Input} (DI) \cite{7140252}. This is because the selected input has the ability to prune the SCK and find at least one incorrect key that is removable from SCK. In addition each time a new DI is found, the SCK search space for function $F_{DI}$ should be updated. This could be achieved by forcing the $F_{DI}$ to check each pair of new keys $K_1$ and $K_2$ against all previously founds DIs. A Complete-DI-set is a set of DI inputs that reduces the SCK to the \textit{Set of Valid Keys} (SVK). SCK reduces to SVK when we no longer can find a DI using the updated $F_{DI}$. At this point if a key is valid across the Complete-DI-Set, it is the correct key for all other inputs~\cite{7140252}.

\begin{figure*}[tb]
    \centering
    \includegraphics[width=0.95\textwidth]{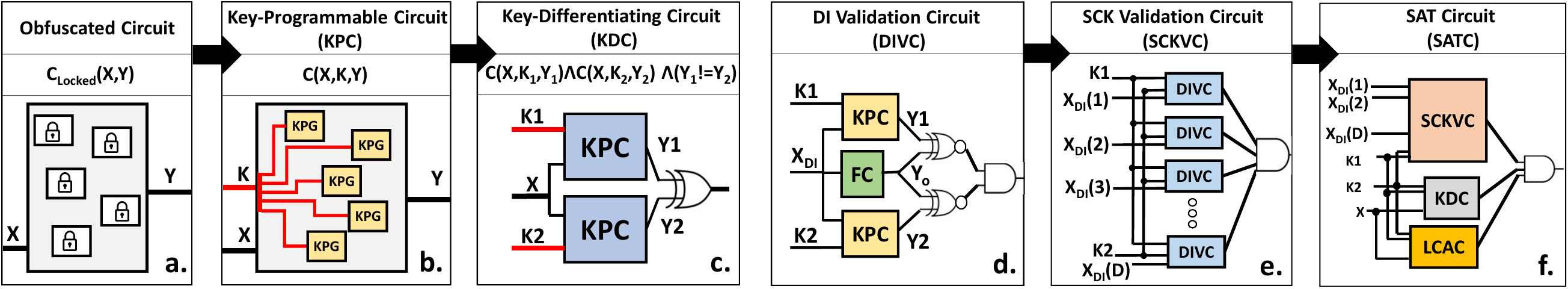}
    \caption{(a) Transforming an obfuscated circuit to (b) Key-Programmable Circuit and (c) Key-Differentiating Circuit. (d) DIVC circuit for validating that two input keys produce the correct output with respect to a previously discovered DI. (e) SCKVC circuit for validating that both input keys are in SCK set and produce the correct output for all previously discovered DIs. (f) SATC circuit for finding a new DI.}
    \label{fig:kpc_satc}
    \vspace{-8pt}
\end{figure*}

In this paper, as suggested in Fig. \ref{fig:kpc_satc}.b, a reverse-engineered netlist, where all obfuscated cells are replaced with KPG cells, is denoted by \textit{Key-Programmable Circuit} (KPC). To build the $F_{DI}$, two copies of the KPC are used, their non-key inputs (X) are tied together, and their outputs are XORed. This circuit produces logic 1 when the output of two instantiated KPCs for the same input X but different keys $K1$ and $K_2$ are different. This circuit, as suggested in Fig. \ref{fig:kpc_satc}.c is denoted as \textit{Key-Differentiating Circuit} (KDC).

The candidate keys in the SCK are capable of producing the correct output for all DIs that have previously been discovered and tested on the KPC circuit. In order to test the keys for one DI, the circuit in Fig. \ref{fig:kpc_satc}.d is instantiated. In this figure, FC is the working copy of the chip, and its output is used for testing the correctness of both KPCs for a given DI and two key values. This circuit is denoted as \textit{DI-Validation Circuit} (DIVC). To test the keys for all DIs, as illustrated in Fig. \ref{fig:kpc_satc}.e, the DIVC circuit is duplicated $D$ times, with $D$ being the number of current DIs tested, and the output of all DIVC circuits ANDed together. The resulting circuit is a validation circuit for SCK set denoted as SCKVC.

If two keys $K_1$ and $K_2$ produce the correct output for all previously tested DIs (SCKVC evaluates to true), but produce different results for a new input $X_{test}$, then $X_{test}$ is a DI that further prunes the SCK. This, as illustrated in Fig. \ref{fig:kpc_satc}.f, could be tested by using an AND gate at the output of SCKVC and KDC circuits. The resulting circuit forms a SAT solvable circuit denoted by SATC. When SATC evaluates to true, the KDC has tested a pair of keys $K_1$ and $K_2$ that produce two different results for an input $X_{test}$, and SCKVC circuit has confirmed that both $K_1$ and $K_2$ belong to SCK set. Hence, the input $X_{test}$ is yet another DI. Each time a new DI is found, the SCKVC should be updated by adding yet another DIVC circuit for testing the newly discovered DI. This process is continued until SAT solver no longer finds a solution to the final SAT circuit. In this case, any key remaining in the SCK set is a correct key for the circuit. On the SAT solver side, every time the SAT solver is executed, it learns a new set of conflict clauses. It is essential to store the learned clauses and use them in the next invocation of the SAT solver to prevent SAT solver from re-learning these clauses. Hence, as illustrated in Fig. \ref{fig:kpc_satc}.f a \textit{Learned-Clause Avoidance Circuit} (LCAC) is added to the SATC to check for the occurrence of learned conflict clauses.  


\section{SAT Attack}
The SAT attack, as illustrated in Algorithm~\ref{SAT_algoritm}, follows the SATC construction process explained in section \ref{SATC_construction}. In the first iteration, the SCKVC circuit does not contain any logic, since there is no previously tested DI. Hence, it is set to 1 (true). The KDC circuit is simply built based on its definition by using the equation in Fig. \ref{fig:kpc_satc}.c. The SATC circuit is constructed by using an ANDing the KDC and SCKVC circuits. $SAT_F$ function is a call to SAT solver. Considering the to-be-assigned variables in SATC circuit are X, $K_1$ and $K_2$, the SAT solvers return an assignment to these variables and a list of \textit{conflict clauses} (CC) learned during SAT execution. $SAT_F$ return UNSAT if no such assignment exists. The while loop is controlled by the return status of the SAT solver. In every pass through the while loop, a new DI is found. Hence, the SATC circuit should be modified (lines 7-10). The parts of SATC circuit that is updated are the SCKVC and LCAC. After finding each DI, an additional DIVC is added to SCKVC to validate the keys generated in the next invocation of SAT solver with respect to the newly found DI. In addition, the newly learned CCs are added to LCAC. The $C_F$ is a call to the functional circuit that returns the correct output for each newly found DI. Finally, the SATC circuit is formulated at line 10 for the next invocation of SAT solver. 

\begin{algorithm}
\caption{SAT Attack on Obfuscated Circuits \label{SAT_algoritm}}
\begin{algorithmic}[1]
\scriptsize
\State $KDC = C(X,K_1,Y_1) \wedge C(X,K_2,Y_2) \wedge (Y_1 \ne Y_2)$;
\State $SCKVC = 1$;
\State $SATC = KDC \wedge SCKVC$
\State $LCAC = 1$

\While {$((X_{DI},K_1,K_2,CC)\leftarrow SAT_F(SATC)=T)$} 
    \State $Y_f \leftarrow C_F(X_{DI})$; 
    \State $DIVC = C(X_{DI},K_1,Y_f) \wedge C(X_{DI},K_2,Y_f)$;
    \State $SCKVC = SCKVC \wedge DIVC$;
    \State $LCAC=LCAC \wedge CC$   
    \State $SATC = KDC \wedge SCKVC \wedge LCAC$;
\EndWhile

\State $KeyGenCircuit = SCKVC \wedge (K_1 = K_2)$
\State $Key \leftarrow SAT_F (KeyGenCircuit)$

\end{algorithmic}
\end{algorithm}

\begin{figure}[ht]
\centering
    \includegraphics[width=0.9\columnwidth]{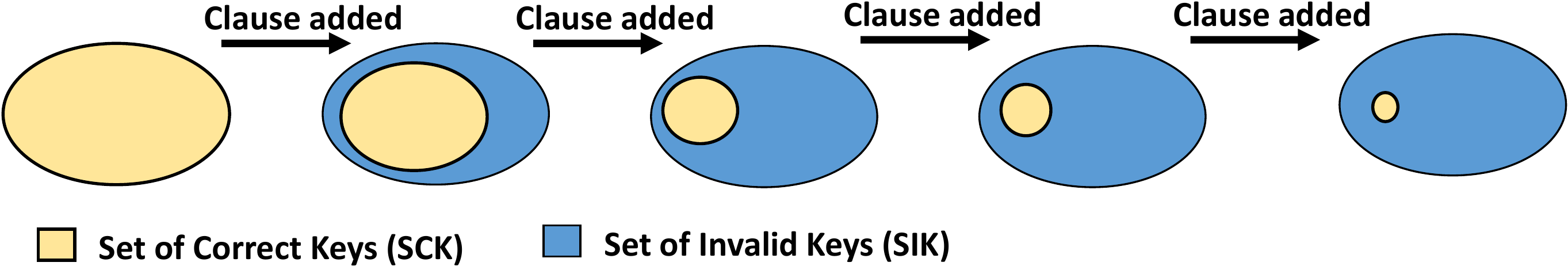}
    \caption{SCK set reduces in each pass through the while loop in Algorithm 1 as a new DI is discovered and is added to SATC circuit.}
    \label{fig:keys_space1}
\end{figure}

The while loop is executed until no other DI is found. At this point, any key in the SCK set is a correct key. To obtain a correct key, the DIVC circuit is modified to take a single key denoted as KeyGenCircuit. Hence, KeyGenCircuit has input $K$, and its output is valid if $K$ satisfy all previous constraints imposed by previously found DIs. A simple call to a SAT solver at this point returns a correct key assignment. If the SAT solver does not return a valid key, it means the obfuscation, locking, or camouflaging technique is invalid. Note that the SAT attack in each iteration, as explained in Algorithm \ref{SAT_algoritm} and illustrated in Fig. \ref{fig:keys_space1}, reduces the SCK by constraining the SATC with new clauses added to the SCKVC and LCAC. But it does not explicitly check to find the keys in SCK.

\section{Benchmarking SAT Solvers' Strengths in Defeating Obfuscation Schemes}
We study the strength of six classes of SAT solvers in defeating 5 previously proposed obfuscation schemes. Our SAT solvers, obfuscation schemes, and the benchmarking platform is described next:

\subsection{SAT Solvers Used in This Work}
MiniSat \cite{een2003extensible}: is developed as a modifiable SAT solver with conflict-driven backtracking, watched literals and dynamic variable ordering. Most of the later SAT solvers are a modified version of this solver. It could be used as a baseline for evaluating the effectiveness of added features in other solvers for obfuscated circuit benchmarks.

Glucose \cite{audemard2015glucose}: is an extension of MiniSat code with a special focus on removing useless clauses as soon as possible and a new restart scheme. It uses the idea of Literal Block Distance (LBD) to estimate the quality of learned clauses. Other SAT solvers incorporated its restart policies. This solver adapts itself according to four predefined outlier benchmark characteristics, but in none of the tested benchmarks, default strategy has changed.

Lingeling \cite{biere2013lingeling}: is based on the idea of interleaving search and pre-processing. It uses various techniques to reduce the search space. Binary and ternary clauses are stored separately from large clauses. Large clauses are kept using literal stacks and references to them are simplified from pointers to stack position. Binary and ternary clauses are kept in occurrence lists. Occurrence lists are defined using stacks and are referenced by stack position. It also uses a modified version of restart mechanism used in Glucose. Number of variables and clauses are also monitored during execution and the number of learned clauses is controlled using their variance.

Maple(MiniSat/Glucose) \cite{liang2016learning}: uses a new branching heuristic in place of Variable State Independent Decaying Sum (VSIDS) called Learning Branching Heuristic (LRB). Two variants of MapleSat are MapleMiniSat and MapleGlucose, respectively based on MiniSat and Glucose. MapleGlucose uses LRB for 2500 seconds of the execution, and then switches to VSIDS. In MapleMiniSat, VSIDS is replaced with LRB.

CryptoMiniSat \cite{soos2010cryptominisat}: is a SAT solver that compiled from SatELite, PrecoSat, Glucose and MiniSat features. It has special mechanisms for XOR clause handling and separates watch lists for binary clauses. It can detect distinct sub-problems in clause list and try to solve them with sub-solvers.

\subsection{Studied Obfuscation Techniques}
A random obfuscation scheme was proposed by Roy et al. in \cite{5604160}. In this scheme, which is one of the earliest work on obfuscation, the XOR/XNOR gates are randomly inserted in the netlist. We refer to this obfuscation as rnd. A major weakness of this scheme was the ability of an attacker to sensitize the circuit, by application of carefully selected inputs, and to propagate the obfuscation keys to the primary outputs of the circuit. Rajendran et al. \cite{rajendran2012security} proposed a more sophisticated obfuscation mechanism to avoid such sensitization attacks by preventing insertion of isolated and mutable key-gates. We refer to this scheme as dac12. An important metric in logic obfuscation is increasing the output corruption when a wrong key is used. Rajendran et al. \cite{6616532} proposed an obfuscation method that uses fault propagation analysis to maximize Hamming distance between correct and incorrect outputs when attacker applies a wrong key. They proposed two variants of their obfuscation technique based on using XOR and MUX gates. We refer to these obfuscation schemes as toc13xor and toc13mux. Wires with low controllability are susceptible to Trojan insertion. To obfuscate the degree of controllability of wires in a netlist, in \cite{6873671} Dupuis et al. tried to minimize the wires with low controllability. This was achieved by inserting AND/OR gates attempting to balance the signal probabilities. We refer to this obfuscation method as iolts14. 

\subsection{Benchmarking Platform}
For benchmarking of selected SAT solvers, we used a farm of 20 Dell Latitude-7010 desktops equipped with Intel Core-i5 processor and 8GB of RAM. For fair comparison, and to reduce the impact of of the operating system background processes, we dedicated one machine to each SAT solver at a time, and installed Ubuntu Server 16.04.3 LTS operating system in shell mode. We used the ISCAS-85 and MCNC benchmark suites in our study and obfuscated each benchmark with \{$1\%, 2\%, 3\%, 5\%, 10\%$ \& $25\% $\} area overhead. To account for run-to-run variation in performance, we ran the SAT solver 15 times for each obfuscated benchmark.

\section{Results}
\begin{figure}[ht]
\centering
    \vspace{-10pt}
    \includegraphics[width=0.9\columnwidth]{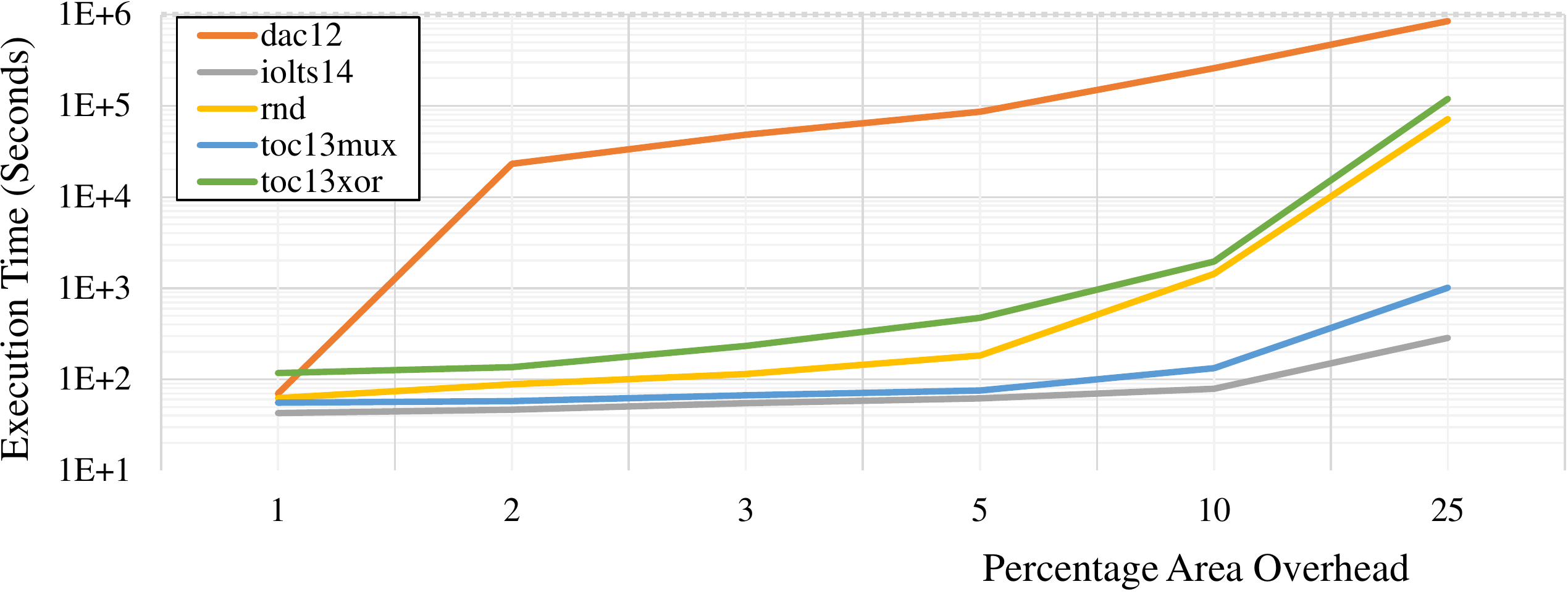}
    \caption{The difficulty of investigated obfuscation solutions across all SAT solvers. The execution time reported is the sum of the execution times of all SAT solvers for finding the key at each reported obfuscation overhead percentage point.}
    \label{fig:obfuscation_difficulty}
\end{figure}

Fig. \ref{fig:obfuscation_difficulty} illustrates the difficulty of defeating each obfuscation method across all SAT solvers. To generate this graph, the execution times for finding the keys to all obfuscated benchmarks are added together at each obfuscation overhead percentage point. The figure illustrates that the complexity of benchmarks obfuscated by dac12 is considerably higher than that for all other investigated obfuscation schemes. We should also note that the time needed for obfuscating a design using the dac12 methodology is considerably longer than the time required by earlier obfuscation methods. The simulation results confirm that increasing the controllability of internal signals, as done in iolts14, or increasing the output corruption, as implemented in toc13, significantly reduces the strength of obfuscation scheme against SAT attacks. Hence, obfuscation schemes that produce the lowest possible output corruption, or reduce the controllability of internal signals pose a harder problem for SAT solvers. However, please note that the aforementioned options for making the obfuscation problem harder for SAT solvers is completely against the reasons why these obfuscation schemes were introduced in the first place (high corruption for higher protection, and high controllability for Trojan prevention). 

For the rest of this section, we separate the discussion of dac12 and the other investigated benchmarks as, depending on the percentage area overhead used for obfuscation, they represent two groups of low-to-mid and mid-to-high complexity SAT problems. Note that we have not used the SAT-hard obfuscation schemes such as SARLock \cite{7495588} for two reasons. First, they are prone to a simpler SPS attack for detection and removal of key-forming-cones. Second, to study the effectiveness of SAT solvers, we deliberately chose to work with medium to semi-difficult problems that are still solvable by SAT solvers in a reasonable time, so that the execution time of SAT solvers is a measure of their efficiency. Otherwise, if the operation of SAT solvers is reduced to brute-force attacks by working on a non-SAT or extreme SAT-hard problems, the execution time of all SAT solvers will be similar, as they will run until they are timed out, or they exhaustively try all possible inputs, thus reducing the SAT solver to a brute-force depth first search solver.

\begin{figure}[ht]
\centering
    \includegraphics[width=0.9\columnwidth]{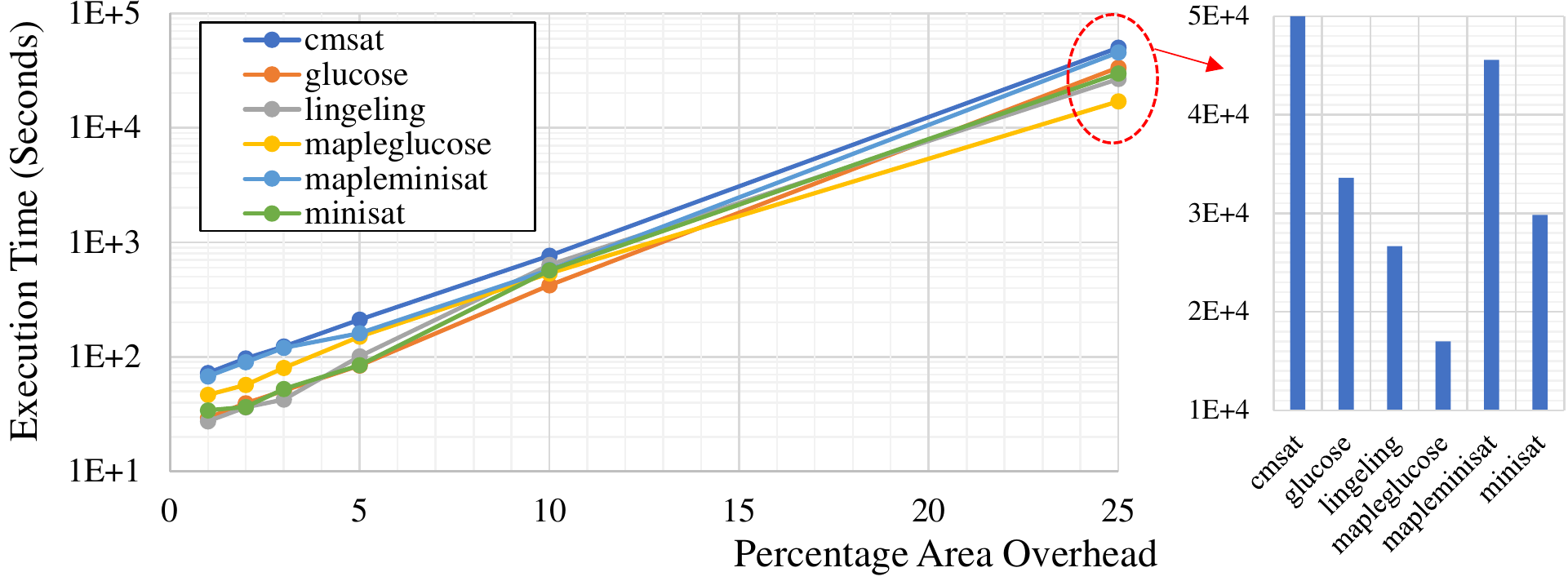}
    \caption{Total execution time of each SAT solver for finding the correct key for all benchmarks and all obfuscation schemes except dac12 as a measure of solver strength for low-to-mid complexity problems.}
    \label{fig:AllSAT}
\end{figure}

\begin{figure}[ht]
\centering
    \vspace{-10pt}
    \includegraphics[width=0.9\columnwidth]{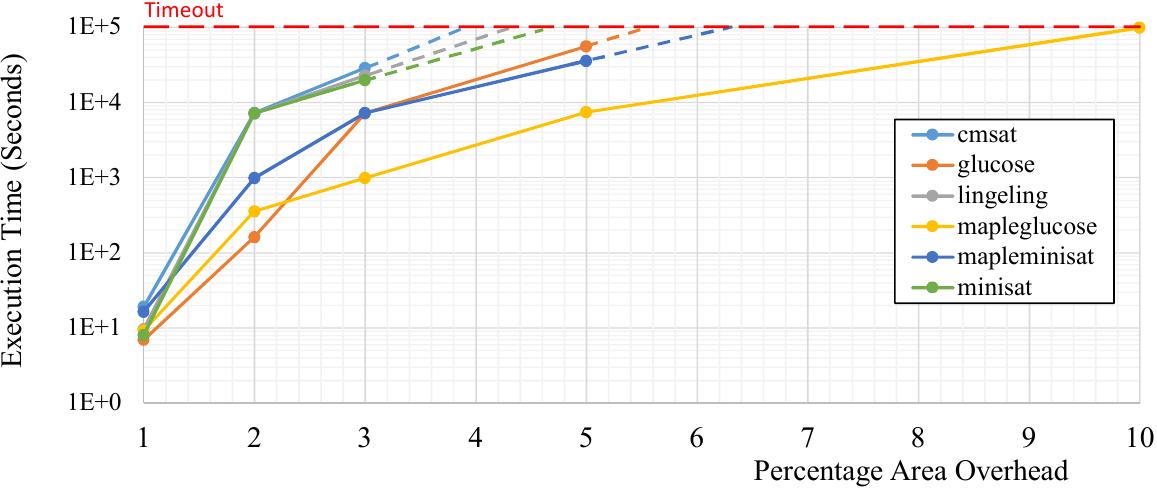}
    \caption{The execution time of SAT solvers for finding the correct key for all dac12 obfuscated benchmarks as a measure of solver strength for mid-to-high complexity problems.}
    \label{fig:DAC12SAT}
\end{figure}

Fig. \ref{fig:AllSAT} (left) illustrates the ability of SAT solvers in defeating the obfuscation scheme across all low-complexity obfuscation schemes (all obfuscation methods except dac12). As illustrated in this figure, the relationship between execution time and area overhead is exponential. However, note that the execution time grows at a very different pace for different SAT solvers, leading to a significant difference in runtime at higher obfuscation percentages. This is illustrated in Fig.~\ref{fig:AllSAT} (right), where the runtime of SAT solvers under study, benchmarked at 25\%, is plotted. As shown in this figure, MapleGlucose, although not the best SAT solver at smaller percentages, outperforms all other solvers by a considerable margin for high percentages, to the point that its runtime is about 3x smaller than that of CryptoMiniSat. Fig. \ref{fig:DAC12SAT} illustrates the ability of investigated SAT solvers to find the key for the netlists obfuscated using dac12. As the obfuscation complexity increases, the runtime of SAT solvers widely varies. In this experiment, a 24-hour limit was imposed on the SAT solvers to break the obfuscated benchmarks. MapleGlucose outperformed all other SAT solvers in this experiment. 

\begin{figure}[ht]
\centering
    \includegraphics[width=0.9\columnwidth]{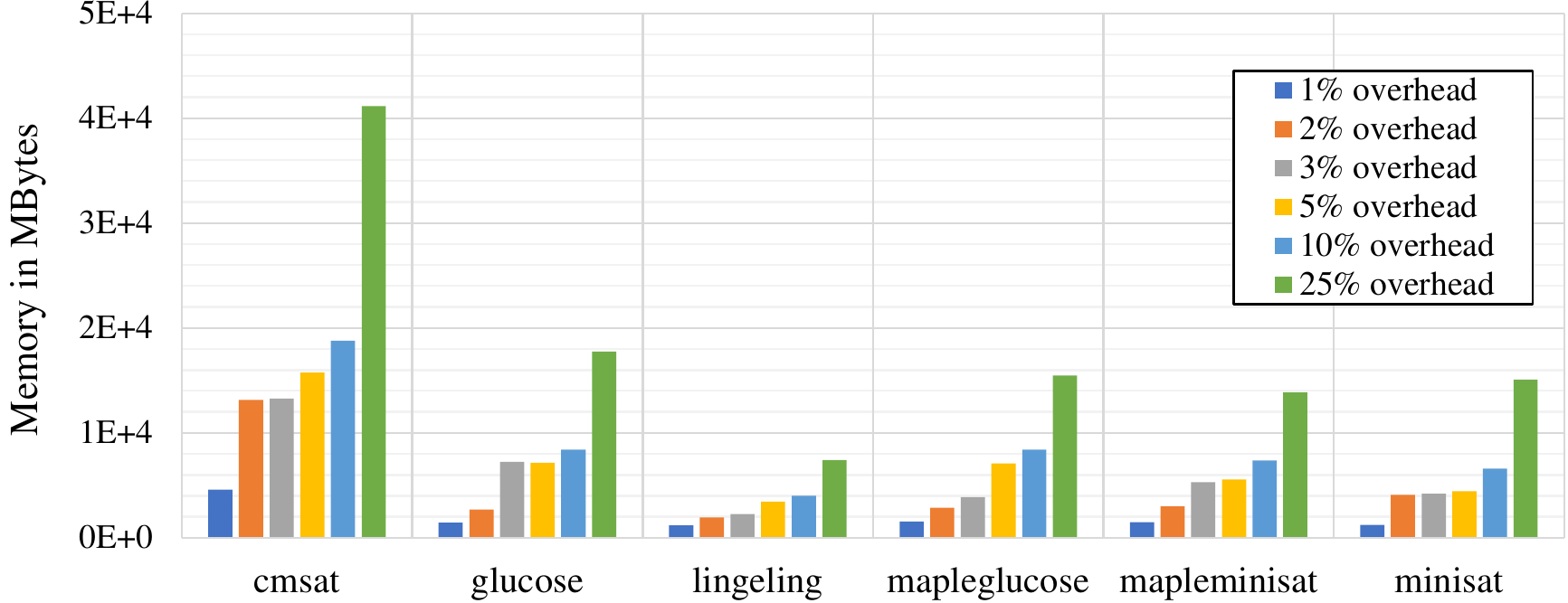}
    \caption{Memory usage of each SAT solver across all benchmarks for each obfuscation percentage point as a measure of solver efficiency.}
    \label{fig:memory}
    \vspace{-10pt}
\end{figure}

Fig. \ref{fig:memory} illustrates the peak of the memory usage for each SAT solver across all benchmarks and at each obfuscation area overhead percentage point. As illustrated in this figure, Lingeling has the lowest memory requirements across all SAT solvers. Hence, it is the most efficient solver in a memory constrained environment, or when the size and percentage of obfuscation considerably increases. As illustrated in Fig. \ref{fig:AllSAT}, Lingeling is also the fastest solver at small obfuscation percentages. At the same time, the CryptoMiniSat is the most memory demanding solver across all obfuscation overheads.

\begin{figure}[ht]
\centering
    \vspace{-10pt}
    \includegraphics[width=0.9\columnwidth]{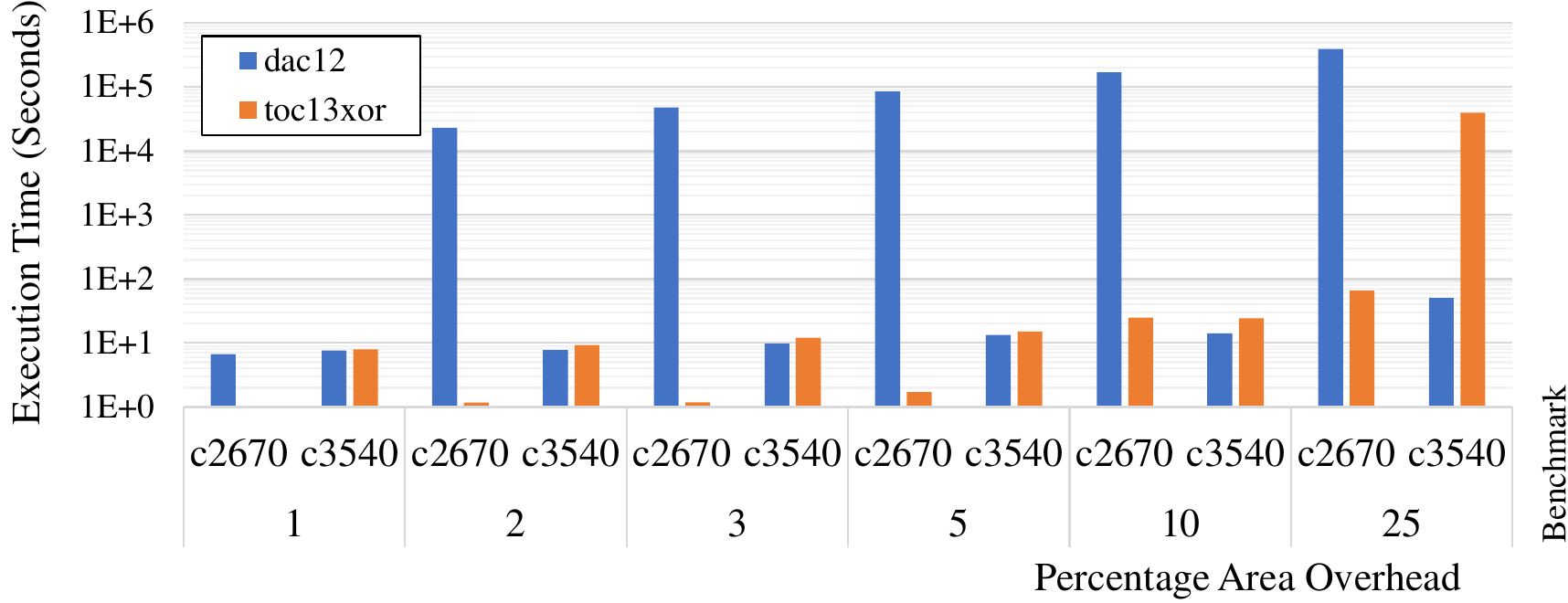}
    \caption{Execution time for deobfuscating c2670 and c3540 which are are obfuscated with toc13xor and dac12.}
    \label{fig:toc13xor}
\end{figure}

In our study, on average, dac12 produced the hardest obfuscation problems for all investigated SAT solvers. However, when it comes to individual benchmarks, we found a few exceptions to this finding, which prevented us from generalizing the result. For example, as illustrated in Fig. \ref{fig:toc13xor}, the total execution time of all SAT solvers for finding keys to benchmarks C2670 and C3540 (being a part of the ISCAS-85 benchmark suite) is compared. The toc13xor obfuscation in circuit C3540 produces a much harder problem for SAT solvers across different obfuscation overheads when compared to dac12, whereas in C2670 the behavior is reversed. Hence, the netlist characteristic (number of inputs, number of gates, connectivity, topology, number of outputs) plays a significant role in the strength of the applied obfuscation, suggesting the use of hybrid obfuscation methods to defend various netlists.    

\section{Discussion and Takeaways}
When it comes to finding keys for a k-obfuscated circuit, the choice of the best SAT solver depends on the netlist characteristics (number of inputs, number of gates, connectivity, topology, number of outputs) and the level of difficulty of implemented obfuscation methods and the available resources of the system executing the SAT solver.

Across the studied solvers, Lingeling provides acceptable performance for small k-obfuscation problems and has the lowest overall memory demand. Our study reveals that Lingeling is best suited for attacking small to midsize obfuscation problems, considering its shorter execution time for these problems, or for attacking extremely large obfuscated circuits, due to its memory efficiency in cases when other solvers become memory-bounded and thus useless. The memory efficiency of Lingeling is the result of a special implementation of data references for 64-bit machines with a specialized memory allocator and garbage collector. 

MapleGlucose, a variation of MapleSat, although not as efficient as Lingeling at small to midsize k-obfuscation problems, still provides the acceptably-good performance. However, in terms of runtime, it significantly outperforms other solvers for large and more difficult k-obfuscation problems. Our investigation revealed that the hybrid branching heuristic used in MapleGlucose proved to be its most useful feature for reducing the solver's execution time. The secondary feature that was observed to be helpful in reducing the MapleGlucose execution time is using the restart policies in Glucose solver. MapleGlucose, however, may not be suited for extremely large problems, as it may fail to execute in a memory-bounded environment, as its memory demands grow faster than for Lingeling.

Our study revealed that the CryptoMiniSat has the worst performance for k-obfuscation, both in terms of execution time, and memory efficiency. CryptoMiniSat incorporates many interesting features and has proven to be powerful, especially for problems that could be partitioned and solved by separate solvers, but the added features do not help with the efficiency of the solver to deal with k-obfuscation problems.  

We experimentally observed that, although different SAT solvers' execution times for a given k-obfuscation problems widely vary, their runtime tracks the obfuscation problems' difficulty. Meaning, if a problem is made more challenging for one solver, it becomes more challenging for all solvers. However, such relationship is not linear. This was especially observed in dealing with the dac12 obfuscation method. Meaning, if a k-obfuscation problem is hardened and SAT solver's execution time is doubled, the problem may cause a much higher or much lower increase in the execution time for another solver. Hence, the results of one solver for a given k-obfuscation cannot be generalized across all SAT solvers.

Across various k-obfuscation methods studied in this paper, dac12 proved to be generally the most difficult. The learned conflict clauses for a dac12 k-obfuscated circuit are usually less constraining as they rely on a larger number of literals. This provides us with a hint to design harder obfuscation problems by exploiting the SAT solver's clause learning behavior and  enforcing mechanisms to increase the number of literals in the learned conflict clauses. Such a defense not only reduces the solver's ability to quickly prune the search space but also increases the memory requirements of the solver for keeping longer clauses. This leads to a faster increase in the size of less effective learned clauses and could degrade the solver in two different ways: (1) The solver memory requirement is pushed towards the system memory bound, (2) the solver's ability to shrink the size of learned clauses based on identification of shorter and more effective (more pruning) clauses is reduced. 

\section{Conclusion}
Our investigation revealed that the Glucose and Lingeling solvers are best suited for small to midsize k-obfuscation problems, while MapleGlucose provides the best execution time for large k-obfuscation problems. When dealing with extremely large k-obfuscation problems, Lingeling again becomes the best choice due to its efficient and less memory demanding database implementation. In terms of testing the hardness of k-obfuscation methods, especially for mid-to-hard size problems, we observed that the increase in the k-obfuscation difficulty affects the runtime of each solver quite differently. Hence, although the increase in difficulty could be verified by one SAT solver, a pace of the increase in difficulty is dependent on the choice of a SAT solver and the results from one solver cannot be generalized. Finally, from a defender's perspective, the results of this benchmarking study suggest that targeting the clause-learning process by means of k-obfuscation, to increase the size of each learned conflict clause, directly affects the effectiveness of SAT solvers in pruning the search space and is a possible promising area for further investigation.


\ifCLASSOPTIONcaptionsoff
  \newpage
\fi

\vspace{-5pt}
\newcommand{\BIBdecl}{\setlength{\itemsep}{0 em}}
\bibliographystyle{IEEEtran}
\bibliography{IEEEabrv,preprint}

\end{document}